\UseRawInputEncoding
\documentclass[preprintnumbers, pra, showpacs ,floatfix,twocolumn,
preprintnumbers, letterpaper, superscriptaddress]{revtex4-1}
\usepackage{amsfonts}
\usepackage{amsmath}
\usepackage{amssymb,epsf}
\usepackage{latexsym}
\usepackage{graphicx,epsfig}
\usepackage{amssymb}
\usepackage{subfigure}
\usepackage[colorlinks=true,citecolor=blue,linkcolor=blue,urlcolor=black]{hyperref}
\usepackage{epstopdf}
\usepackage{color,soul}

\def\be{\begin{equation}}
\def\ee{\end{equation}}
\def\ba{\begin{eqnarray}}
\def\ea{\end{eqnarray}}
\def\la{\langle}
\def\ra{\rangle}

\begin{document}
\title{ٍFoliated order parameter in a fracton phase transition}

\author{Mohammad Hossein Zarei}
\email{mzarei92@shirazu.ac.ir}
\affiliation{Department of Physics, School of Science, Shiraz University, Shiraz 71946-84795, Iran}

\author{Mohammad Nobakht}
\email{mqisnobakht@gmail.com}
\affiliation{Department of Physics, School of Science, Shiraz University, Shiraz 71946-84795, Iran}

\begin{abstract}
Finding suitable indicators for characterizing quantum phase transitions plays an important role in understanding different phases of matter. It is especially important for fracton phases where a combination of topology and fractionalization leads to exotic features not seen in other known quantum phases. In this paper, we consider the above problem by studying phase transition in the X-cube model in the presence of a non-linear perturbation. Using an analysis of the ground state fidelity and identifying a discontinuity in the global entanglement, we show there is a first-order quantum phase transition from a type I fracton phase with a highly entangled nature to a magnetized phase. Accordingly, we conclude that the global entanglement, as a measure of the total quantum correlations in the ground state, can well capture certain features of fracton phase transitions. Then, we introduce a non-local order parameter in the form of a foliated operator which can characterize the above phase transition. We particularly show that such an order parameter has a geometric nature which captures specific differences of fracton phases with topological phases. Our study is specifically based on a well-known dual mapping to the classical plaquette Ising model where it shows the importance of such dualities in studying different quantum phases of matter.
\end{abstract}
\maketitle
\section{Introduction}
Identifying different phases of matter is one of the most important achievements of condensed matter physics during the past century \citep{goldenfeld1992lectures}.
 However, it seems there are still physical models with richer structures than what we know by now. In particular, while the symmetry breaking theory of Landau explains all locally ordered phases of matter, it fails to explain phases with non-local order such as topological phases \citep{wen1989vacuum,22,wen}. Such a different order leads to amazing consequences such as robust degeneracy and exotic excitations with important applications in quantum information processing
\citep{Kitaev2003,15,robustness,robustness2,zarei21}. On the other hand, while physicists are still looking for a comprehensive mechanism for characterizing topological phases, it has been shown that a combination of fractionalization and topological order leads to different quantum phases in three dimensions with interesting properties namely fracton phases \citep{nandkishore2019fractons,sa,haah2011local,chamon2005quantum,bravyi2011topological,bravyi2013quantum,yoshida2013exotic}.

In spite of  similarities with topological phases, there are important challenges with fracton phases. In particular, fracton phases are three-dimensional (3D) models \citep{shirley2018fracton} whose ground state degeneracy exponentially increases by the system size
\citep{vijay2015new,castelnovo2012topological}. Furthermore, there is an interesting physics with excitations where they can be only a fraction of a mobile particle
\citep{shirley2019fractional}. In other words, while fractons are immobile particles, composites of them form topological excitations that can move in lower-dimensional subsystems \cite{ shirley2019foliated}. Such a property causes some challenges for realizing suitable field theory in low energy regime of fracton phases
\citep{slagle2017quantum,q1,q2,q3}. Beside theoretical attractions in different fields \citep{q5,q6,q7,q8,q9}, the lack of mobility for fractons has important applications in quantum information processing. In particular, fracton phases are important candidate for quantum error correcting codes where fracton nature of excitations is in favor of error correction which is even more efficient than conventional topological codes at finite temperature \citep{terhal2015quantum,brown2016quantum}.

Among different models for fracton phases, the X-cube model, which is defined on a 3D cubic lattice, is a simple and important one \citep{vijay2016fracton,slagle2018x}. It is categorized as a type I fracton phase \cite{devakul2018correlation} where topological excitations are not immobile, but they can move in planes or lines of the lattice, and consequently they are named planons or lineons. This model has a foliated structure with an interesting duality with 3D classical plaquette Ising model which is a spin model with planar symmetries \citep{vijay2016fracton}. Regarding such a duality, a foliated field theory has also been proposed for the low energy physics of the X-cube model \citep{slagle2021foliated}.

An important approach for understanding fracton phases such as the X-cube model is to study quantum phase transitions out of such phases \cite{poon2021quantum,extra1}. An important issue is to consider different indicators which can characterize a fracton phase transition. In particular, it is important to find out how the behavior of a certain indicator at the critical point reflects the certain properties of the fracton models. For example, recently the X-cube model in presence of a magnetic field has been studied \citep{muhlhauser2020quantum} where a first-order quantum phase transition is identified by a discontinuity in the first derivation of the ground state energy. The first-order nature of phase transition in this model reveals its difference with topological phases such as toric code model which shows a second-order phase transition in presence of a parallel magnetic field
\citep{trebst2007breakdown,zarei2019ising,reiss2019quantum}.

Here we are going to take a step towards better understanding fracton phases by studying different indicators for a fracton phase transition. In particular, our focus here is on properties of the ground state wave function instead of the ground state energy where we have two important goals. First, since fracton phases have wave functions with a non-local nature similar to topological phases, we expect that there is a non-local order parameter which is able to characterize phase transition in fracton models. In particular, since order parameters contain all important information about their corresponding phases, it is important to find how the specific features of the fracton phases reveal in a possible non-local order parameter \cite{zarei2019ising}. Second, it is known that phase transitions can be characterized by measures of quantum information theory such as fidelity and entanglement \citep{hamma2008entanglement,gu2010fidelity,yang2008fidelity,marzolino2017fisher,carollo2018uhlmann}. Since such measures are directly related to properties of the wave function of the ground state, their study plays an important role in understanding the nature of different quantum phases \cite{gl,montakhab,elahe}. Accordingly, it is important to study how different measures of the entanglement as well as ground state fidelity can characterize a fracton phase transition.

In this regard, we consider a fracton phase transition in the X-cube model in the presence a non-linear perturbation. We use a well-known duality between Calderbank-Shor-Steane (CSS) codes and classical spin models to find the exact ground state of the model \cite{zarei2020classical,zaremon}. Accordingly, we analytically calculate the ground state fidelity for the perturbed X-cube model and show that it is mapped to the heat capacity in the 3D plaquette Ising model. Using the well-known singularity in the heat capacity of the 3D plaquette Ising model, we characterize the phase transition from a fracton phase to a magnetized one in the quantum model. Then, we study the global entanglement as a measure of total quantum correlation in the ground state. We show that the global entanglement shows a discontinuity at the transition point which reveals the first-order nature of the fracton phase transition as well as highly entangled nature of the fracton phase. Finally, we introduce a foliated order parameter with a non-local nature and show how certain features of the fracton phase reflect in the geometrical properties of such an order parameter.

The structure of the paper is as follows: In Sec.(\ref{sec1}) we introduce the perturbed X-cube model and use mapping between the ground state fidelity and the heat capacity in the 3D plaquette Ising model to identify the quantum phase transition point in the quantum model. In Sec.(\ref{sec2}), we characterize the first-order nature of the quantum phase transition by identifying a discontinuity in the global entanglement. Finally, in Sec.(\ref{sec3}), we propose the foliated order parameter as a suitable indicator to characterize certain features of the fracton phase.

\section{X-cube model in presence of a non-linear perturbation}\label{sec1}
\begin{figure}[t]
\centering
\includegraphics[width=5cm,height=5cm,angle=0]{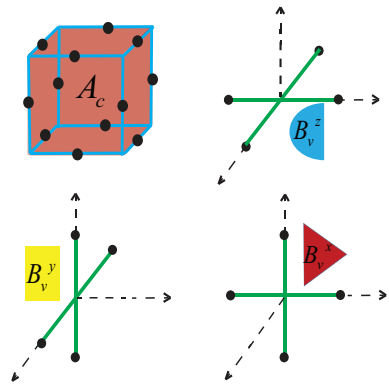}
\caption{(Color online) Stabilizer operators of the X-cube model. Qubits are denoted by black circles. Corresponding to each cube $c$, an $A_c$ stabilizer is defined as a product of $\sigma^x$ Pauli operators around the cube $c$. Corresponding three different planes crossing a vertex $v$, there are also three types of vertex operators of $B_v ^z$, $B_v ^y$ and $B_v ^x$ denoted by the blue semicircle, the yellow rectangle and the red triangle, respectively. }\label{stab}
\end{figure}

X-cube model is defined on an $L\times L \times L$ 3D lattice where qubits live in the edges of the lattice
\citep{vijay2016fracton}.
 There are $X$-type stabilizers corresponding to cubic cells of the lattice as well as three kinds of $Z$-type stabilizers corresponding to vertices of the lattice in the following form, see Fig. (\ref{stab}):
$$A_c =\prod_{i\in c} \sigma^x _i~~,~~B_{v}^{x}=\prod_{i\in v_{x}}\sigma^z_i$$
\begin{equation}
B_{v}^{y}=\prod_{i\in v_{y}}\sigma^z _i~~,~~B_{v}^{z}=\prod_{i\in v_{z}}\sigma^z _i ,
\end{equation}
where $i\in c$ refers to qubits belonging to the cell of $c$ and $i\in v_{x}$, $i\in v_{y}$ and $i\in v_{z}$ refer to four qubits incoming to the vertex $v$ and living in a plane with normal vectors $\hat{x}$, $\hat{y}$ and $\hat{z}$, respectively. Accordingly, the Hamiltonian of the X-cube model is given by:
\begin{equation}
H_x =-\sum_c A_c -\sum_v (B_{v}^{x}+B_{v}^{y}+B_{v}^{z})
\end{equation}

Since the $X$-type and the $Z$-type stabilizers commute with each other, one of the ground states of the above Hamiltonian is simply found in the following form, up to a normalization factor:
\begin{equation}
|G_{x}\ra =\prod_c (1+A_c)|0\ra^{\otimes N},
\end{equation}
where $1$ refers to the Identity operator and $N$ refers to the number of qubits of the lattice. Under periodic boundary condition, there are also many constraints between stabilizers which lead to a large degeneracy in the ground state. In particular, it is shown that there are $6L-3$ constraints between stabilizers, and it leads to a $2^{6L-3}$-fold degeneracy in the ground state \cite{vijay2016fracton}.

\begin{figure}[t]
\centering
\includegraphics[width=8cm,height=4cm,angle=0]{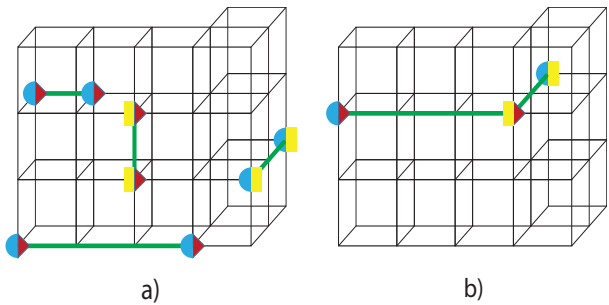}
\caption{(Color online) a) a $\sigma^x$ operator applied in an edge qubit, denoted by green (thick) color, does not commute with two types of vertex stabilizers living at two endpoints of the corresponding edge. The resulting excitations live in the corresponding vertices and are denoted by semicircles, rectangles or triangles depending on the type of stabilizer that is violated. Applying a chain of $\sigma^x$ operators along a specific direction leads to moving the corresponding excitations. b) The excitations can only move in one dimension. When one wants to move in another direction, it leaves an additional excitation in the corner of the path. }\label{verex}
\end{figure}

On the other hand, one can generate topological excitations of the above Hamiltonian simply by applying $\sigma^x$ or $\sigma^z$ operators to qubits of the lattice
\citep{vijay2016fracton}. In particular, $\sigma^x$ operator which is applied to a qubit $i$, does not commute with four vertex operators of $B$, which qubit $i$ belongs to. As shown in Fig.(\ref{verex}-a), it leads to excitations living in the corresponding vertices. In particular, if we apply $\sigma^x$ operators along a chain of qubits, a couple of excitations will move in the direction of the chain. In other words, it is impossible to have a single excitation moving in the lattice, but two excitations should be regarded as a moving particle. Therefore, a single excitation is a fraction of a moving particle and, consequently, is called a fracton. Moreover, topological excitations can only move along a line of the lattice and are called lineons because, by applying $\sigma^x$ operator in a qubit out of the above line, it leaves an additional excitation in the corner of the path, see Fig.(\ref{verex}-b).

\begin{figure}[t]
\includegraphics[width=80mm,scale=1.5]{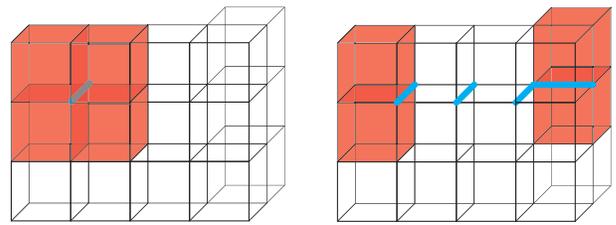}
\caption{(Color online) a $\sigma^z$ operator applied in an edge qubit, denoted by blue (thick) color, does not commute with four cubic stabilizers living in four cubes around the corresponding edge. The resulting excitations are denoted by pink cubes. They can also move in a two-dimensional plane by applying a suitable chain of $\sigma^z$ operators. }\label{cubeex}
\end{figure}

Another kind of excitation is generated by applying a $\sigma^z$ operator on an edge qubit $i$. It does not commute with four cell operators which $i$ belongs to. Therefore, we have four excitations where a couple of them can move by applying $\sigma^z$ operators along a chain of qubits. However, these topological excitations are free to move in any direction in a plane, and therefore they are called planons, see Fig.(\ref{cubeex}).

Fracton phases similar to topological phases have important topological properties. For example, in the ground state of the X-cube model, one can apply a chain of $\sigma^x$ operators on qubits along a closed string
\citep{haah2011local}. Such an operation leads to another ground state of the model which is topologically distinguishable from the first one. Other degenerate ground states can be generated in the same way and all of them are topologically distinguishable. Furthermore, the degree of degeneracy in the X-cube model depends on the boundary conditions similar to topological phases. However, fractionalization of excitations leads to some important differences between fracton phases and topological phases. In particular, while topological order is independent of the geometry of the underlying lattice, some properties of the fracton phases depend on the geometry
\citep{slagle2018x,shirley2018fracton}.

Now, we are ready to introduce a non-linear perturbation to the X-cube model in the form of:
\begin{equation}\label{hj}
H=H_x + \sum_c e^{-\beta \sum_{i\in c} \sigma^z _i},
\end{equation}
where $\beta$ refers to a perturbation parameter which is non-linearly coupled to the X-cube model. The above Hamiltonian is an example of a general CSS-Hamiltonian introduced in \citep{zarei2020classical} where there is a dual mapping between such CSS-Hamiltonians and classical spin models. According to such a mapping, the ground state of the Hamiltonian (\ref{hj}) is in the following simple form:
\begin{equation}\label{wq}
|G_{x} (\beta)\rangle =\frac{1}{\sqrt{\mathcal{Z}}}e^{\frac{\beta}{2}\sum_i \sigma^z_i}|G_x\rangle ,
\end{equation}
where $\mathcal{Z}$ in the normalization factor is mapped to the partition function of a classical spin model\cite{zaremon}. We particularly show that $\mathcal{Z}$ is in fact the partition function of a plaquette Ising model defined on a 3D cubic lattice with classical spins living in the vertices. Corresponding to each plaquette of each cube of such lattice, there is a four-body interaction with a classical Hamiltonian in the form of $H_{cl}=-\sum_{p} s_i s_j s_k s_l$ where $s_i$'s are classical spins and $p$ refers to a square plaquette which spins $i, j, l $ and $k$ belong to.

\begin{figure}[t]
\includegraphics[width=80mm,scale=1.5]{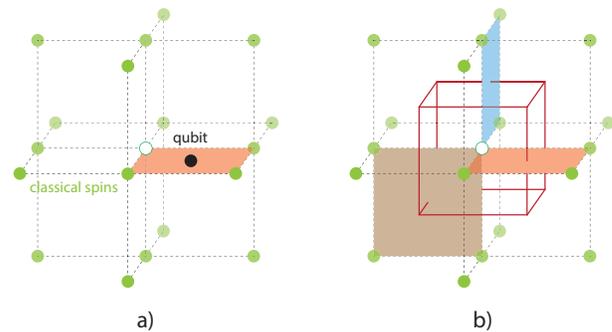}
\caption{(Color online) a) Classical spins in the plaquette Ising model live in the vertices, denoted by green (light) circles. Corresponding to each plaquette, there is a four body interaction term. We insert one qubit, denoted by a black (dark) circle, in the center of each plaquette. b) It is a particular spin configuration where the white spin is set to $-1$ and other spins are set to $+1$. The white spin is surrounded by a cubic in the dual lattice in which all qubits living in the edges are $|1\ra$. }\label{essential}
\end{figure}

In order to prove the above mapping, we insert one quantum bit in the center of each plaquette corresponding to the interaction term of $s_i s_j s_k s_l$ in the sense that if $s_i s_j s_k s_l=1(-1)$, we set the corresponding qubit to $|0\ra$($|1\ra$), see Fig. (\ref{essential}-a). Then, we plot a 3D cubic dual lattice where the above qubits live on the edges of that lattice. In other words, the original classical spins live in the centers of cubes in the dual lattice. Then as shown in Fig.(\ref{essential}-b), consider a spin configuration where one of the classical spins, which is denoted by a white circle and is surrounded by a cube of the dual lattice, is $-1$ and the other spins are $+1$. In such a configuration, all interaction terms including the white circle have the value of $-1$ and therefore, all edge qubits living in the corresponding dual cube are $|1\rangle$. In this regard, each classical spin configuration in the initial lattice corresponds to a cubic structure of qubits of $|1\ra$ in a sea of qubits of $|0\ra$ in the dual lattice. Furthermore, we emphasize that there are different spin configurations which lead to the same qubic structure. In particular, if we flip classical spins living in each plane of the lattice, the classical Hamiltonian will be invariant. Therefore, since there are $3L$ planes in the lattice, there are $2^{3L}$ spin configurations that lead to the same cubic structure.

 Now let us come back to the ground state of the X-cube model. In particular, we note that in the $|G_{x}\rangle=\prod_{c}(1+A_c)|0\rangle^{\otimes N}$, one can span $\prod_{c}(1+A_c)$ in the form of a summation of all $X$-type stabilizers constructed by product of $A_c$s. Since each $A_c$ corresponds to one cubic of the lattice, a product of $A_c$ can be represented by a combination of cubes of the lattice. In this regard, when we apply $\prod_{c}(1+A_c)$ to the product state of $|0\ra ^{\otimes N}$ and since $\sigma^x |0\ra=|1\ra$, it leads to a superposition of different cubic structures constructed by $|1\ra$ in a sea of $|0\ra$s. They are exactly the same structures that we saw in the dual of the plaquette Ising model. In other words, there is a duality between spin configurations in the 3D plaquette Ising model and cubic structures in the ground state of the X-cube model in the sense that corresponding to each cubic structure in the X-cube model, there are $2^{3L}$ spin configurations of the 3D plaquette Ising model.

Finally, note that since the value of Ising interactions of $s_i s_j s_k s_l$ in the plaquette Ising model is $1$ or $-1$, the energy of an arbitrary spin configuration is equal to $E=2Q-N$ where $Q$ is the number of interaction terms with the value of $-1$ and $N$ is the total number of plaquettes. On the other hand, consider Eq.(\ref{wq}) where we have applied $e^{\frac{\beta}{2}\sum_i \sigma^z _i}$ to the state of $|G_{x}\rangle$. When $e^{\frac{\beta}{2}\sum_i \sigma^z _i}$ is applied to each cubic structure in the $|G_{x}\rangle$, we will have a superposition of the cubic structures with amplitudes in the form of $e^{\frac{\beta}{2}(N-2Q)}$. Therefore, it is concluded that the amplitude of each cubic structure is equal to the root of the Boltzmann weight of the corresponding spin configuration in the 3D plaquette Ising model if we map the perturbation parameter $\beta$ to the inverse temperature $1/k_B T$ where $k_B$ is the Boltzmann constant. In this regard, it is clear that the normalization factor of $\mathcal{Z}$ should be the same as the partition function of the plaquette Ising model up to a factor of $2^{3L}$. In other words, the Boltzmann weights of the 3D plaquette Ising model have been interestingly encoded in the amplitudes of the ground state of the perturbed X-cube model in the sense that each Pauli operator of $\sigma^z$ in the quantum model has been mapped a four-spin interaction in the dual classical model.

 Now, we show that the above important mapping leads to an interesting connection between a thermal phase transition in the 3D plaquette Ising model and a quantum phase transition in the perturbed X-cube model. To this end, we consider the ground state fidelity in the perturbed X-cube model in the form of $F=\la G(\beta)|G(\beta +d\beta)\ra$ \cite{gu2010fidelity}. As it is shown in \cite{zarei2020classical}, this quantity is mapped to the heat capacity of the dual plaquette Ising model by the following relation:
\begin{equation}\label{er}
F=1-\frac{C_v}{8\beta^2}d\beta^2 +O(d\beta^3),
\end{equation}
where $O(d\beta^3)$ refers to terms in higher degrees when we expand $F$ in terms of $d\beta$ and $C_v$ refers to the heat capacity of the 3D plaquette Ising model.

\begin{figure}[t]
\includegraphics[width=80mm,scale=1.5]{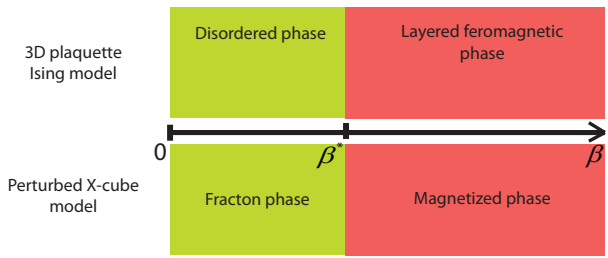}
\caption{(Color online) A schematic of duality between phase structures of the 3D plaquette Ising model and the perturbed X-cube model. $\beta$ is the inverse temperature $1/k_B T$ (perturbation parameter) in the 3D plaquette Ising model (in the perturbed X-cube model) where, by increasing $\beta$, a phase transition from a disordered phase (fracton phase) to a layered ferromagnetic phase (magnetized phase) occurs. The first order phase transition happens at $\beta^* =0.551$}\label{schemat}
\end{figure}

On the other hand, it is known that there is a strong first-order phase transition in the 3D plaquette Ising model \cite{espriu1997evidence} where internal energy shows discontinuity at a transition temperature \citep{espriu1997evidence,e3,dimopoulos2002slow}. Therefore, the heat capacity for 3D plaquette Ising model shows a singularity (divergence) at the transition point of $T^*$ . In this regard, Eq.(\ref{er}) implies that there is also a singularity in the ground state fidelity at a $\beta^* =1/k_B T^*$ where a quantum phase transition in the perturbed X-cube model occurs. According to recent studies on 3D plaquette Ising model \cite{sd}, the transition point is $\beta^* =0.551$. In particular, note that at $\beta=0$ the quantum model is a pure $X$-cube model, and we have a fracton phase. It corresponds to $T\rightarrow \infty$ in the 3D plaquette Ising model where there is a disordered phase. On the other hand, at $\beta \rightarrow \infty$ the ground state (\ref{wq}) becomes a simple magnetized state of $|000...0\ra$, and it corresponds to $T\rightarrow 0$ in the 3D plaquette Ising model where there is a layered ferromagnetic order due to planar up-down symmetries of the model, see Fig.(\ref{schemat}).

Since the phase transition in the 3D plaquette Ising model is of the first-order, we know that the corresponding quantum phase transition is also of the first-order. It is in agreement with another study in \cite{muhlhauser2020quantum} where authors study the ground state energy for the X-cube model in presence of a magnetic field and show that there is a discontinuity in its first-order derivative. As it is explained in the above paper, the first-order nature of the fracton phase transition is a result of the immobility of fractons. Here, we are going to do a different analysis based on studying the global entanglement \cite{meyer} in the ground state wave function. Since the global entanglement is a measure of total quantum correlation, it is also an important task to consider how certain features of the fracton phase such as immobility of fractons can reflect in the behavior of the global entanglement at the transition point.

\section{Discontinuity of the global entanglement}\label{sec2}
Global entanglement has a value between zero and one and is simply defined as the average of the linear entropy of one-qubit density matrices in the following form:
\begin{equation}\label{ui}
GE=2[1-\frac{1}{N}\sum_p tr(\rho_p ^2)],
\end{equation}
where $\rho_p$ refers to the reduced density matrix of a single qubit of $p$.

It has been shown that the global entanglement can well characterize the order of a quantum phase transition \cite{gl}. In this section, we calculate this quantity for the ground state of the perturbed X-cube model $|G_{x}(\beta)\rangle$ to show the first-order nature of the corresponding quantum phase transition.

We now find the one-qubit density matrix on a computational basis by tracing out other qubits in $|G_{x}(\beta)\rangle \langle G_{x}(\beta)|$:
\begin{equation}\label{po}
\rho_p =\frac{1}{\mathcal{Z}}\sum_{\{\mu_n =0,1 \}; n \neq p}\langle \mu_1 \mu_2 ...\mu_N| G_{x}(\beta)\rangle \langle G_{x}(\beta)| \mu_1 \mu_2 ...\mu_N \rangle,
\end{equation}
where $\rho_p$ is a $2\times 2$ matrix with matrix elements of $\la \mu_p |\rho_p |\mu'_p\rangle$. Regarding Eq.(\ref{po}), the above matrix elements can be written in the following form:
$$\la \mu_p |\rho_p |\mu'_p\rangle =\frac{1}{\mathcal{Z}}\sum_{\{\mu_n =0,1 \};n \neq p}$$
  \begin{equation}\label{pi}
\langle \mu_1 \mu_2 ...\mu_p ...\mu_N| G_{x}(\beta)\rangle \langle G_{x}(\beta)| \mu_1 \mu_2 ...\mu'_p ...\mu_N \rangle
\end{equation}
Since that $|G_{x}(\beta)\rangle$ is a superposition of different cubic structures constructed by $|1\ra$'s in the sea of $|0\ra$'s, it is clear that the inner product of $\langle \mu_1 \mu_2 ...\mu_p ...\mu_N| G_{x}(\beta)\rangle$ in the above equation will be equal to zero if $\langle \mu_1 \mu_2 ...\mu_p ...\mu_N|$ does not correspond to a cubic structure. In this regard, it is concluded that non-diagonal elements of $\rho_p$ must be zero, because it is impossible that both $\langle \mu_1 \mu_2 ...\mu_p ...\mu_N|$ and $\langle \mu_1 \mu_2 ...\mu'_p ...\mu_N|$ in Eq.(\ref{pi}) correspond to the same cubic structure while $\mu_p \neq \mu'_p$. On the other hand, for diagonal elements where $\mu_p =\mu'_p$, we only need to consider $\langle \mu_1 \mu_2 ...\mu_p ...\mu_N|$'s corresponding to cubic structures. In particular, $\mu_p =0$ corresponds to cubic structures that do not cross the qubit of $p$ and  $\mu_p =1$ corresponds to cubic structures that cross the qubit of $p$. In this regard, we have the diagonal elements of the one-qubit density matrix in the following form:
$$\rho_p^{00} =\frac{1}{\mathcal{Z}}\sum_{cubic-s_0}|\langle \mu_1 \mu_2 ...0 ...\mu_N| G_{x}(\beta)\rangle|^2  $$
\begin{equation}\label{rt}
\rho_p^{11} =\frac{1}{\mathcal{Z}}\sum_{cubic-s_1}|\langle \mu_1 \mu_2 ...1 ...\mu_N| G_{x}(\beta)\rangle|^2,
\end{equation}
where $cubic-s_{1(0)}$ refers to cubic structures crossing (not crossing) the qubit of $p$.

 Next, we remind that $| G_{x}(\beta)\rangle$ was also a superposition of cubic structures with different amplitudes which were equal to the root of the Boltzmann weights corresponding to different spin configurations in the 3D plaquette Ising model. In this regard, it is concluded that the $|\langle \mu_1 \mu_2 ...1(0) ...\mu_N| G_{x}(\beta)\rangle|^2$ is equal to the Boltzmann weight of a particular spin configuration corresponding to a cubic structure that crosses (does not cross) the qubit of $p$. As we explained in the previous section, in cubic structures corresponding to each spin configuration, the state of a qubit is $|1\ra (|0\ra)$ if the corresponding interaction term of $E_p =s_i s_j s_k s_l =-1(+1)$. Therefore,  $\sum_{cubic-s_1}$ ($\sum_{cubic-s_0}$) is equal to a summation of all Boltzmann weights corresponding to spin configurations with the interaction term of $E_p =-1(+1)$. Therefore, regarding the factor of $\frac{1}{\mathcal{Z}}$, we conclude that the diagonal elements of $\rho_p$ in Eq.(\ref{rt}) are equal to probabilities that the interaction term of $E_p$ takes the value of +1 or -1 at a temperature of $T$:
\begin{equation}\label{rt}
\rho_p^{11}=W_-  ~~,~~\rho_p^{00}=W_+
\end{equation}
In particular, the statistical expectation value of the interaction term of $E_p$ is equal to $\bar{E_p}=W_+ -W_-$.  On the other hand, since $W_+ +W_- =1$, it is simply concluded that:
\begin{equation}\label{ryt}
W_+ =\frac{1+\bar{E_p} }{2} ~~,~~W_- =\frac{1-\bar{E_p} }{2}
\end{equation}
In this regard, the diagonal terms of the one-qubit density matrix of the perturbed X-cube model are mapped to the expectation value of the corresponding interaction term in the 3D plaquette Ising model. By replacing (\ref{ryt}) in Eq.(\ref{ui}), the global entanglement is written in the following form:
\begin{equation}\label{pl}
GE=2(1-\frac{1}{N}\sum_p \frac{1+\bar{E_p} ^2}{2})
\end{equation}
Furthermore, according to the uniform interaction pattern of the plaquette Ising model with periodic boundary condition, it is concluded that $\bar{E_p}$ should be the same for all interaction terms. Therefore, one can conclude that $\bar{E_p}=\frac{U(N,T)}{N}$ where $U(N,T)$ refers to internal energy of the model. By replacing in (\ref{pl}), the global entanglement is simply written in terms of internal energy in the form of:
\begin{equation}
GE(\beta)=1-u(T)^2 ,
\end{equation}
 where $u$ is internal energy per plaquette which has a value between $0$ to $-1$. In particular, notice that in the low temperature ordered phase, we have $u=-1$ and consequently, $GE=0$. It confirms that the ground state of the perturbed X-cube model at large values of $\beta$ is a magnetized (product) state without any entanglement. On the other hand, in the high temperature disordered phase, we have $u=0$ and therefore, $GE=1$. It confirms also that the ground state of the perturbed X-cube model at small values of $\beta$ is in the fracton phase with a highly entangled nature.

Furthermore, it is important to find the internal energy for arbitrary temperature in the plaquette Ising model to see how the global entanglement shows a transition from $1$ to $0$ at the transition point $\beta^*$. Fortunately, in \cite{espriu1997evidence} authors have simulated the plaquette Ising model and plotted the internal energy in terms of $\beta=1/k_B T$ for finite size lattices. In particular, they have shown that there is a discontinuity in the internal energy at the transition point $\beta^*$. Therefore, it is concluded that the global entanglement also shows a discontinuity at the same transition point. It clearly confirms that the above quantum phase transition in the perturbed X-cube model has a first order nature. Discontinuity of the total quantum correlation in the ground state reveals also certain features of the fracton phases. In particular, in \cite{elahe} authors have considered the global entanglement for a topological phase transition in a perturbed toric code model where the global entanglement shows a continuous (second-order) transition. In this regard, discontinuity of the global entanglement in the perturbed X-cube model should be a result of the immobility of fractons which distinguishes fracton phases from conventional topological phases.
%

 \section{A foliated order parameter}\label{sec3}
As we explained in the previous section, it is important to note that any quantum phase transition can be explained by a mechanism in terms of the behavior of excitations in the model under consideration. In particular, the first-order nature of phase transition in the perturbed X-cube model is indeed a result of limitations in mobility of the excitations. In particular, one can compare this model with the perturbed toric code model \cite{robustness} with a second-order quantum phase transition. Interestingly, the second-order phase transition in the toric code can be characterized by a string (non-local) order parameter
\citep{zarei2019ising}
 which well reveal topological nature of phase transition. In this regard, it is also important to identify a suitable order parameter to characterize the first-order phase transition in the X-cube model and then show that how it reveals fracton properties of the X-cube model.

\begin{figure}[t]
\includegraphics[width=60mm,scale=0.5]{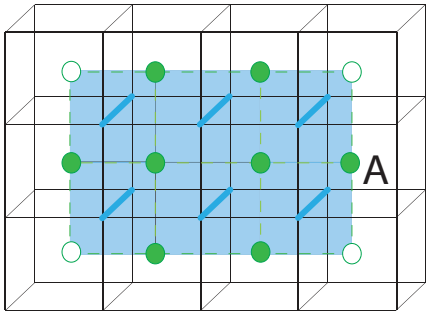}
\caption{(Color online) Qubits belonging to the plane $A$ are denoted by the blue (thick) edges of the cubes. Classical dual spins live in the center of cubes and are denoted by circles. Each $\sigma^z$ operator corresponding to one edge qubit is mapped to an interaction term between four classical spins around that qubit. Therefore, a product of $\sigma^z$ operators belonging to $A$ is mapped to an interaction term between spins living in the corners of the $A$ denoted by white circles. }\label{dual}
\end{figure}

To this end, we introduce a non-local operator in the form of product of $\sigma^z$ operators in the form of $O_A= \prod_{i \in A} \sigma^z _i$ where $A$ refers to an arbitrary 2D membrane living in the plane of the lattice, see Fig.(\ref{dual}), and $i\in A$ refers to all qubits living in the $A$. Moreover, in order to guarantee the non-local nature of $O_A$, we suppose that the size of the $A$ is of the order of the system size. Then, we show that the expectation value of such an operator is mapped to a multi-spin correlation in the plaquette Ising model. To this end, note that we can extend the model Hamiltonian in Eq.(\ref{hj}) in the following form:
\begin{equation}\label{hjo}
H=H_x + \sum_c e^{- \beta \sum_{i\in c} J_i \sigma^z _i},
\end{equation}
where $J_i$ can be different corresponding to different qubits and the model reduces to our original model when $J_i$'s set to $1$.

Similar to our original mapping, the extended model is also mapped to a dual 3D plaquette Ising model in which classical spins live in the center of each cube of the X-cube model. However, the dual classical Hamiltonian is in the form of $H=-J_ p\sum_{p}s_i s_j s_k s_l$ where $J_p$ refers to the coupling constant corresponding to the plaquette of $p$, and it is the same as $J_i$ in Eq.\ref{hjo}. In other words, each $\sigma^z$ operator in the X-cube model is mapped to one corresponding four-spin interaction in the plaquette Ising model. The ground state of the model also has the following form:
\begin{equation}\label{wqe}
|G_x (\beta)\rangle =\frac{1}{\sqrt{\mathcal{Z}}}e^{\frac{\beta}{2}\sum_i J_i \sigma^z _i}|G_x\rangle
\end{equation}
The above form is suitable for calculating the expectation value of the operator of $O_A$ in the following form:
\begin{equation}\label{}
\la O_A \ra =\frac{1}{\mathcal{Z}}\la G_x | (\prod_{i\in A}\sigma^z _i ) e^{\beta \sum_i J_i \sigma^z _i} |G_x \rangle
\end{equation}
To this end, it is enough to replace each $\sigma^z $ operator with $\frac{1}{\beta}\frac{\partial}{\partial J_i}$ and, consequently
\begin{equation}\label{}
\la O_A \ra =\frac{1}{\mathcal{Z}}\prod_{i\in A}(\frac{1}{\beta}\frac{\partial}{\partial J_i})\mathcal{Z}
\end{equation}

On the other hand, since $\mathcal{Z}$ is the partition function of $H=-J_ p\sum_{p}s_i s_j s_k s_l$, it is clear that each term of $\frac{1}{\beta}\frac{\partial}{\partial J_i}$ in the above equation leads to a $s_i s_j s_k s_l$ term. Consequently, $\prod_{i\in A}(\frac{1}{\beta}\frac{\partial}{\partial J_i})$ leads to classical expectation value of a product of $s_i s_j s_k s_l$ terms corresponding to all plaquettes belonging to the membrane $A$, see Fig.(\ref{dual}). However, since $s_i ^2 =1$, it is concluded that the product of  $s_i s_j s_k s_l$ terms belonging to $A$ is equal to a multi-spin interaction in the form of $\prod_{i \in corner}s_i$ where $i\in corner $ refers to all spins living in corners of the membrane $A$, denoted by white circles in the Figure.

According to the above mapping between the expectation value of the non-local operator of $O_A$ in the perturbed X-cube model and the expectation value of the corresponding multi-spin interaction in the plaquette Ising model, we are able to consider the behavior of $\la O_A\ra$ in terms of $\beta$ in the quantum phase transition in our model. In particular, since the size of the $A$ is of the order of the system size, it is concluded that the spins living in four corners of the square membrane of $A$ are very far from each other. On the other hand, it is known that the correlation functions exponentially decay in terms of distance between spins due to the finite correlation length. Therefore, we conclude that the expectation value of the four-spin interaction must be the same as the expectation value of a single spin power to $4$, i.e $\la s_1 s_2 s_3 s_4 \ra =\la s_1 \ra \la s_2 \ra \la s_3 \ra \la s_4 \ra$. In this regard, it is enough to determine the expectation value of a single spin in the plaquette Ising model $\langle s_i \rangle$. In other words, while $\langle s_i \rangle$ is a local parameter in the plaquette Ising model, its behavior is exactly mapped to the behavior of the non-local parameter of $\la O_A\ra$ in the perturbed X-cube model.

\begin{figure}[t]
\centering
\includegraphics[width=80mm,scale=1.5]{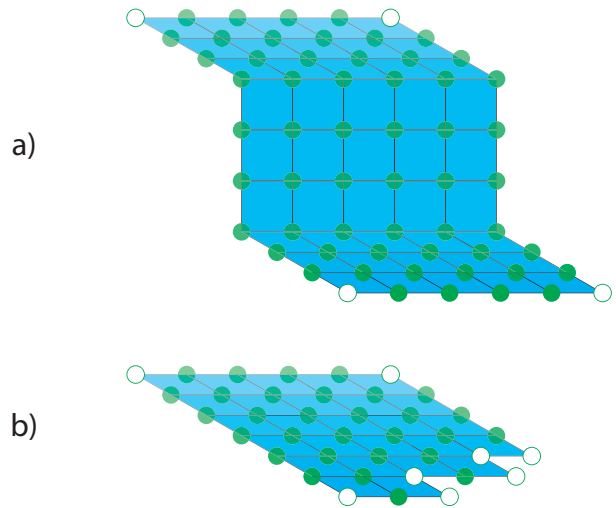}
\caption{(Color online) Classical spins are denoted by circles and, qubits live in the centers of plaquettes. a) A membrane operator formed by combination of different planes is mapped to a multi-spin interaction between four white spins living in the corners of the membrane. b) A local deformation in the plane leads to additional corners and the corresponding operator is mapped to a multi spin interaction including all corners.}\label{plaquette}
\end{figure}

On the other hand, note that the 3D plaquette Ising model has planar symmetries where the magnetization of each 2D plane of the lattice has an up-down symmetry. It means that in the low temperature phase, the system has a layered ferromagnetic order where the magnetization of each plane is determined independently. In this regard, the magnetization per spin for a particular plane is a suitable order parameter to characterize the thermal phase transition. In particular, note that the magnetization per spin for the whole system is always equal to zero and can not be an order parameter. However, as it has been shown in \cite{espriu1997evidence,e3}, one can consider a fixed boundary condition for the plaquette Ising model in a sense that all planes are forced to have the same magnetization. Under such a condition, the magnetization per spin for each plane is the same as the magnetization per spin for the whole lattice and it can be derived by a suitable simulation. In particular, in \cite{espriu1997evidence} authors shows that  the magnetization shows a discontinuous behavior at the transition point which reveals the first-order nature of the phase transition in the 3D plaquette Ising model.

 Next, it is enough to notice that, in the thermodynamic limit, $\langle s_i \rangle$ is also equal to the magnetization per spin for a plane which $i$ belongs to. Consequently, according to mapping to the non-local operator of $O_A$, there must be also a discontinuous behavior for $\la O_A \ra$ in terms of the $\beta$ and it clearly identifies the first-order nature of the corresponding quantum phase transition in the perturbed X-cube model. Furthermore, as shown in fig.(\ref{plaquette}-a), we are also able to choose various membrane operators as an order parameter. The important point is that the product of four-spin interactions corresponding to all plaquettes of such a membrane is again equal to a spin interaction between the corners. It is in fact a result of the underlying foliated structure in the X-cube model and therefore, we call the $\la O_A \ra$ a \textit{foliated} order parameter.

It is also important to compare the above order parameter in the X-cube model with string order parameter for toric code model introduced in \cite{zarei2019ising}. In the toric code model, the string order parameter is mapped to a two-spin interaction term in a simple Ising model corresponding to two spins living at two endpoints of the string. In particular, this mapping does not depend on the form of the string in the sense that if one does an arbitrary local deformation in the corresponding string, it is again mapped to the same two-spin interaction. This property is expected due to topological order in the toric code model because topological properties are invariant in terms of local deformation. However, the situation is different with the X-cube model. In particular, if one does a local deformation in the membrane $A$, it leads to new corners in the above membrane. Consequently, the initial spin interaction converts to a new spin interaction in which more spins contribute, see Fig.(\ref{plaquette}-b). In particular, if the distance between spins is small, the corresponding spin interaction can not be mapped to the magnetization. This interesting property of the foliated order parameter in the X-cube model confirms new study done in \cite{slagle2018x} where authors propose that fracton phases are geometric phases and not topological phase. In other words, while topological properties should be independent of the geometry, it seems that when the geometry is changed by the local deformations, it leads to differences in the behavior of some important quantities in the fracton phases.

\section{Conclusion}
It seems that there are still many problems with fracton phases of matter because of exotic constraints on mobility of excitations. In this regard, it is an important task to study how certain features of the fracton phases reflect on microscopic as well as macroscopic behaviors of fracton models. Among different approaches, studying quantum phase transitions out of fracton phases plays an important role. In particular, here we considered some different properties of the ground state of a perturbed X-cube model where we identified a first-order phase transition signed by a singularity in the ground state fidelity. Then, from a microscopic view, we studied the total quantum correlation of the ground state by calculating the global entanglement where we showed that it shows a discontinuity at the transition point. Furthermore and from a macroscopic view, we introduced a foliated order parameter with a non-local and geometrical nature where, unlike conventional topological phases, there is also an important role for the geometry. It in particular, sheds light on distinctive aspects of fracton phases with topological phases and can help to better understand the nature of these exotic phases of matter. Finally, we would like to emphasize that our results here were based on a classical-quantum mapping which is an example of a general duality mapping between classical spin models and quantum CSS codes \cite{zarei2020classical}. Therefore, our paper paves the way for future studies for using such dualities for considering other fracton CSS codes including type-II fracton phases.

\section*{Acknowledgement}
We would like to thank A. Ramezanpour for fruitful discussions.

\end{document}